\documentclass[twocolumn,floatfix,showpacs,aps,prd]{revtex4}

\usepackage{graphicx}
\usepackage{dcolumn}
\usepackage{amsmath}
\usepackage{units}

\input{babarsym}

     \def\dtoll {\ensuremath{D^0\to\ell^+\ell^-}\ }
     \def\dtopipi {\ensuremath{D^0\to\pi^+\pi^-}\ }
     \def\dtokpi {\ensuremath{D^0\to K^-\pi^+}\ }
     \def\dtoee {\ensuremath{D^0\to e^+e^-}\ }
     \def\dtomumu {\ensuremath{D^0\to \mu^+\mu^-}\ }
     \def\dtoemu {\ensuremath{D^0\to e^\pm\mu^\mp}\ }
     \def\dzero {\ensuremath{D^0}\ }
     \def\mdzero {\ensuremath{m(D^0)}\ }
     \def\deltam {\ensuremath{\Delta m}\ }
     \def\mumu  {\ensuremath{\mu^+\mu^-}\ }
     \def\ee    {\ensuremath{e^+e^-}\ }
     \def\emu   {\ensuremath{e^\pm \mu^\mp}\ }
     \def\pipi  {\ensuremath{\pi^+\pi^-}\ }
     
     \def\bbbar {\ensuremath{B\bar{B}}\ }
     \def\coshel{\ensuremath{|\cos \theta_{\rm hel} |}}

     \newcommand{\smtvs}{\rule[-0.15cm]{0cm}{0.5cm}}

\def\figurebox#1#2#3{%
    \def\arg{#3}%
    \ifx\arg\empty
    {\hfill\vbox{\hsize#2\hrule\hbox to #2{\vrule\hfill\vbox to #1{\hsize#2\vfill}\vrule}\hrule}\hfill}%
    \else
    {\hfill\epsfbox{#3}\hfill}%
    \fi}

\long\def\inst#1{\par\nobreak\kern 4pt\nobreak
    {\it #1}\par\vskip 10pt plus 3pt minus 3pt}

\begin{document}

\begin{flushleft}
 BABAR-PUB-12/009 \\
 SLAC-PUB-15093 \\
\end{flushleft}

\title{
  { \Large \bf \boldmath Search for the decay modes \dtoee, \dtomumu, and \dtoemu }
}
%
\author{J.~P.~Lees}
\author{V.~Poireau}
\author{V.~Tisserand}
\affiliation{Laboratoire d'Annecy-le-Vieux de Physique des Particules (LAPP), Universit\'e de Savoie, CNRS/IN2P3,  F-74941 Annecy-Le-Vieux, France}
\author{J.~Garra~Tico}
\author{E.~Grauges}
\affiliation{Universitat de Barcelona, Facultat de Fisica, Departament ECM, E-08028 Barcelona, Spain }
\author{A.~Palano$^{ab}$ }
\affiliation{INFN Sezione di Bari$^{a}$; Dipartimento di Fisica, Universit\`a di Bari$^{b}$, I-70126 Bari, Italy }
\author{G.~Eigen}
\author{B.~Stugu}
\affiliation{University of Bergen, Institute of Physics, N-5007 Bergen, Norway }
\author{D.~N.~Brown}
\author{L.~T.~Kerth}
\author{Yu.~G.~Kolomensky}
\author{G.~Lynch}
\affiliation{Lawrence Berkeley National Laboratory and University of California, Berkeley, California 94720, USA }
% ...
\collaboration{The \babar\ Collaboration}
\noaffiliation

\collaboration{The \babar\ Collaboration}

\date{\today}

\begin{abstract}
   We present searches for the rare decay modes
   \dtoee, \dtomumu, and \dtoemu in continuum $e^+e^- \to c\bar c$
   events recorded by the \babar\ detector in a data sample that
   corresponds to an integrated luminosity of 468~fb$^{-1}$.
   These decays are highly GIM suppressed
   but may be enhanced in several extensions of the Standard Model.
   Our observed event yields are consistent with the expected
   backgrounds.
   An  excess is seen in the \dtomumu channel, although the
   observed yield is consistent with an upward background fluctuation
   at the 5\% level.
   Using the Feldman-Cousins method, we set the
   following 90\% confidence level intervals on the branching
   fractions:
     ${\cal B}(\dtoee)<1.7 \times 10^{-7}$ ,
     ${\cal B}(\dtomumu)$ within $[0.6, 8.1]\times 10^{-7}$ , and
     ${\cal B}(\dtoemu)<3.3 \times 10^{-7}$ .

\end{abstract}

\pacs{
13.20.Fc,11.30.Hv,12.15.Mm,12.60.-i
}

\maketitle

\section{ Introduction }

   In the Standard Model (SM), the flavor-changing neutral current (FCNC) decays
   \dtoll are strongly suppressed by the Glashow-Iliopoulos-Maiani (GIM) mechanism.
   Long-distance processes bring the predicted branching fractions
   up to the order of $10^{-23}$ and $10^{-13}$ for \dtoee and \dtomumu decays,
   respectively~\cite{burdman}.
   These predictions are well below current experimental sensitivities.
   The lepton-flavor violating (LFV) decay \dtoemu is forbidden in the SM.
   Several extensions of the SM predict \dtoll branching fractions that
   are enhanced by several orders of magnitude compared
   with the SM expectations~\cite{burdman}.
   The connection between
   \dtoll and
   $D^0 - \bar{D^0}$ mixing
   in new physics models has also been emphasized~\cite{golowich}.

   We search for \dtoll decays using approximately 468~fb$^{-1}$ of data
   produced by the PEP-II asymmetric-energy $e^+e^-$ collider~\cite{pepii}
   and recorded by the \babar\ detector.
   The center-of-mass energy of the machine was
   at, or 40~MeV below, the $\Upsilon(4S)$ resonance
   for this dataset.
   The \babar\ detector is described in detail elsewhere~\cite{babar-nim}.
   We give a brief summary of the main features below.

   The trajectories
   and decay vertices of long-lived hadrons
   are reconstructed with a 5-layer, double-sided silicon strip
   detector (SVT) and a 40-layer drift chamber (DCH), which are inside a 1.5 T solenoidal
   magnetic field.
   Specific ionization ($dE/dx$) measurements are made by both the
   SVT and the DCH.
   The velocities of charged particles are inferred from the
   measured Cherenkov angle
   of radiation emitted within fused silica bars, located outside the tracking volume and
   detected by an array of phototubes (DIRC).
   The $dE/dx$ and Cherenkov angle measurements are used in particle identification.
   Photon and electron energy, and photon position, are
   measured by a CsI(Tl) crystal
   calorimeter (EMC).
   The steel of the flux return for the solenoidal magnet is instrumented
   with layers of either resistive plate chambers or limited streamer tubes~\cite{menges},
   which are used to identify muons (IFR).

 \section{ Event reconstruction and selection }

   We form \dzero candidates by combining pairs of oppositely charged tracks
   and consider the following final states: \ee, \mumu, \emu, \pipi, and
   $K^-\pi^+$.
   We use the measured \dtopipi yield and the known \dtopipi branching
   fraction to normalize our \dtoll branching fractions.
   We also use the \dtopipi candidates, as well as the \dtokpi candidates,
   to measure the probability of misidentifying a $\pi$ as either
   a $\mu$ or an $e$.
   Combinatorial background is reduced by requiring that the \dzero
   candidate originate from the decay $D^*(2010)^+ \to \dzero \pi^+$~\cite{chrgconj}.
   We select \dzero candidates produced in continuum $e^+e^- \to c \bar c$
   events by requiring that the momentum of the \dzero candidate be
   above 2.4 GeV in the center-of-mass (CM) frame, which is
   close to the kinematic limit for $B \to D^*\pi$, $D^{*+}\to D^0\pi^+$.
   This reduces the combinatorial background from $e^+e^- \to \BB$ events.

   Backgrounds are estimated directly from data control samples.
   Signal \dzero candidates with a reconstructed \dzero mass above 1.9~GeV
   consist of random combinations of tracks.
   We use a sideband region above the signal region in the \dzero mass
  ([1.90, 2.05] GeV) in a wide $\Delta m \equiv m(\dzero \pi^+) - m(D^0)$
   window ([0.141, 0.149] GeV)
   to estimate the amount of combinatorial background.
   The \dzero and $\Delta m$ mass resolutions, measured in the \dtopipi sample,
   are 8.1~MeV and 0.2~MeV, respectively.
   We estimate the number of \dtopipi background events selected
   as \dtoll candidates by scaling the observed \dtopipi yield, with no
   particle identification criteria applied, by the product of pion misidentification
   probabilities and a misidentification correlation factor $G$.
   The misidentification correlation factor $G$ is estimated with the
   \dtokpi data control sample.

   The tracks for the \dzero candidates must have momenta greater than 0.1 GeV
   and have at least 6 hits in the SVT.
   The slow pion track from the $D^{*+} \to \dzero \pi^+$ decay must have at
   least 12 position measurements in the DCH.
   A fit of the $D^{*+} \to \dzero \pi^+; D^0 \to t^+t^-$ decay chain is
   performed where the \dzero tracks $(t)$ are constrained to come from a common
   vertex and the \dzero and slow pion are constrained to form a common vertex
   within the beam interaction region.
   The $\chi^2$ probabilities of the \dzero and $D^*$ vertices from this fit must
   be at least $1$\%.
   The reconstructed \dzero mass \mdzero must be within $[1.65, 2.05]$~GeV and
   the mass difference $\Delta m$ must be within $[0.141, 0.149]$~GeV.
   We subtract a data-Monte-Carlo difference of  $0.91 \pm 0.06$~MeV,
   measured in the \dtopipi sample,
   from the reconstructed \dzero mass in the simulation.

   We use an error-correcting output code (ECOC) algorithm~\cite{ecoc} with
   36 input variables to identify electrons and pions.
   The ECOC combines multiple bootstrap aggregated~\cite{bagging} decision
   tree~\cite{decision-tree}
   binary classifiers trained to separate $e, \pi, K,$ and $p$.
   The most important inputs for electron identification
   are the EMC energy divided by the track momentum,
   several EMC shower shape variables, and the deviation from the expected
   value divided by the measurement uncertainty for the
   Cherenkov angle and $dE/dx$ for the $e, \pi, K,$ and $p$ hypotheses.
   For tracks with momentum greater than 0.5 GeV, the electron
   identification has an efficiency of 95\% for electrons and
   a pion misidentification probability of less than 0.2\%.
   Neutral clusters in the EMC that are consistent with Bremsstrahlung
   radiation are used to correct the momentum and energy of electron
   candidates.
   The efficiency of the pion identification is above 90\% for pions,
   with a kaon misidentification probability below 10\%.

   Muons are identified using a bootstrap aggregated decision tree algorithm with
   30 input variables. Of these, the most important are
   the number and positions of the hits in the IFR, the
   difference between the measured and expected DCH $dE/dx$
   for the muon hypothesis, and the energy deposited in the EMC.
   For tracks with momentum greater than 1~GeV, the muon
   identification has an efficiency of around 60\% for muons, with
   a pion misidentification probability of between 0.5\% and 1.5\%.

   The reconstruction efficiencies for the different
   channels after the above particle identification
   requirements are about 18\% for $e^+e^-$, 9\% for $\mu^+\mu^-$, 13\% for $e^\pm\mu^\mp$,
   and 26\% for $\pi^+\pi^-$.
   The background candidates that remain are either random
   combinations of two leptons (combinatorial background), or
   \dtopipi decays where both pions pass the lepton identification
   criteria (peaking background).
   The \dtopipi background is most important for the \dtomumu channel.

\begin{figure}
      \begin{center}
         \includegraphics[width=0.49\linewidth]{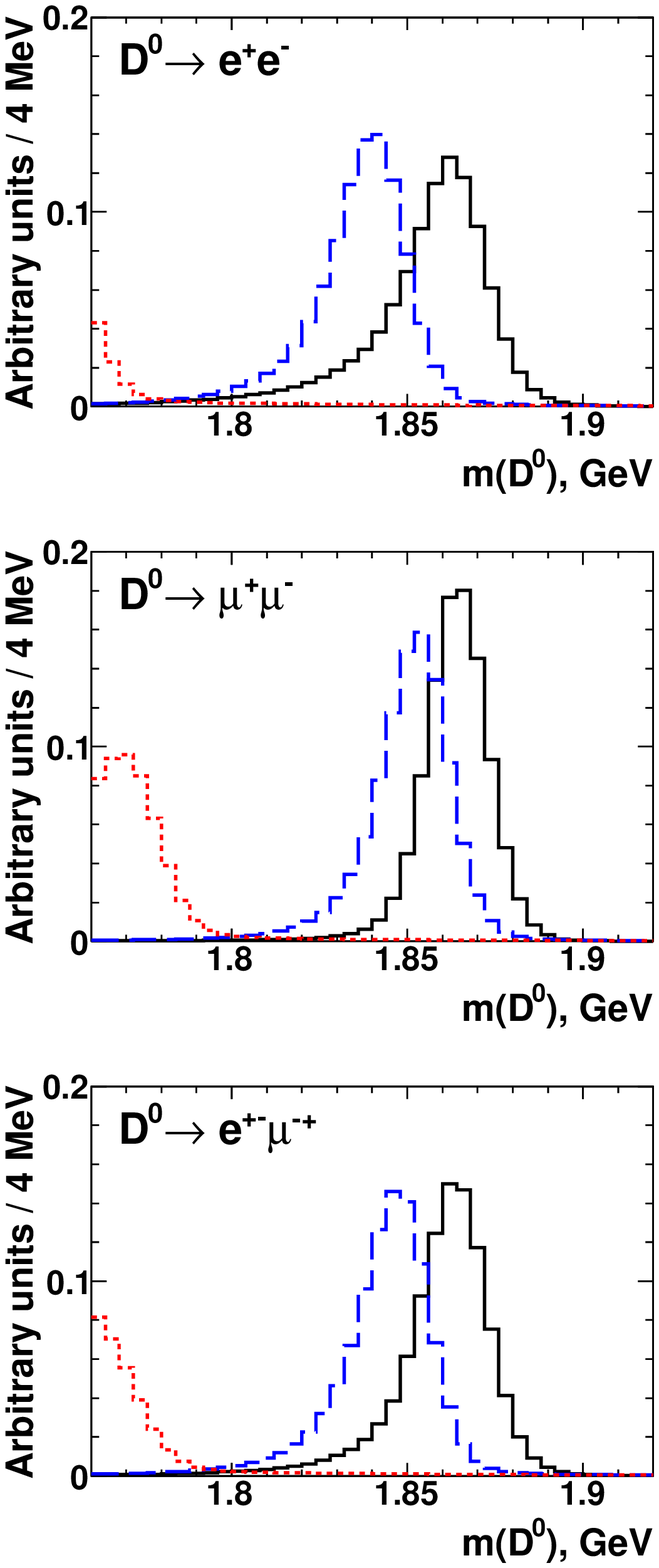}
         \includegraphics[width=0.49\linewidth]{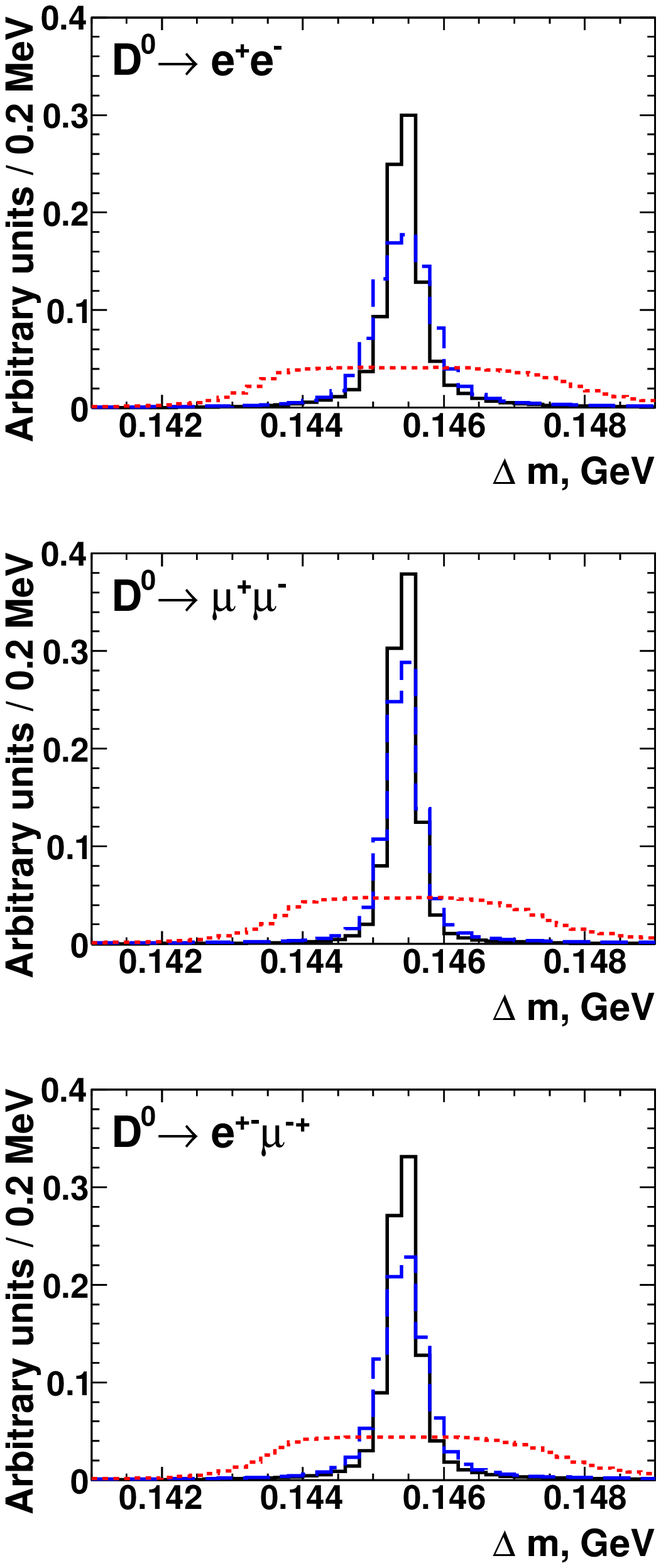}
         \caption{ 
                   Reconstructed \dzero mass (left) and \deltam\ (right) for the three signal channels:
                   \dtoee (top), \dtomumu (middle), and \dtoemu (bottom).
                 The solid (black) histogram is the signal MC, the dashed (blue)
                 histogram is \dtopipi MC reconstructed as \dtoll,
                 and the dotted (red) histogram is \dtokpi MC reconstructed as \dtoll.
                 The \dtoll and \dtopipi distributions have been normalized to unit area.
                 The \dtokpi normalization is arbitrary.
                 }
         \label{fig:d0m}
      \end{center}
   \end{figure}

   Figure~\ref{fig:d0m} shows the reconstructed invariant mass distributions
   from Monte Carlo (MC) simulated samples for the three \dtoll signal channels.
   Also shown are the distributions from \dtopipi reconstructed as
   \dtoll and \dtokpi reconstructed as \dtoll for each signal channel.
   The overlap between the \dtoll and \dtopipi distributions is
   largest for the \dtomumu channel, while the \dtoll and
   \dtokpi distributions are well separated.

   The combinatorial background originates mostly from events with two
   semileptonic $B$ and/or $D$ decays.
   The sample of events selected by the above criteria are dominantly from
   $e^+e^- \to B\bar B$ events, rather than
   events from the $e^+e^- \to q\bar q,\ (q=u,d,s,c)$ continuum.
   We use a linear combination (Fisher discriminant~\cite{fisher}) of the following five
   variables to reduce the combinatorial \bbbar background:
   \begin{itemize}
     \item The measured \dzero flight length divided by its uncertainty.
     \item The value of \coshel, where $\theta_{\rm hel}$ is defined as the angle
           between the momentum of the positively-charged \dzero daughter and
           the boost direction from the lab frame to the \dzero rest frame,
           all in the \dzero rest frame.
     \item The missing transverse momentum with respect to the beam axis.
     \item The ratio of the $2^{\rm nd}$ and $0^{\rm th}$ Fox-Wolfram moments~\cite{r2}.
     \item The \dzero momentum in the CM frame.
   \end{itemize}
   The flight length for combinatorial background is symmetric about zero, while the
   signal has an exponential distribution.
   The \coshel\ distribution is uniform for signal but peaks at zero for
   combinatorial \bbbar background.
   The neutrinos from the semileptonic decays in \bbbar background events
   create missing transverse momentum, while there is none for signal
   events.
   The ratio of Fox-Wolfram moments uses general event-shape
   information to separate \bbbar and continuum $q\bar q$ events.
   Finally, the signal has a broad \dzero CM momentum spectrum that peaks
   at around 3 GeV, while combinatorial background peaks at the minimum
   allowed value of 2.4 GeV.

\begin{figure}
      \begin{center}
         \includegraphics[width=\linewidth]{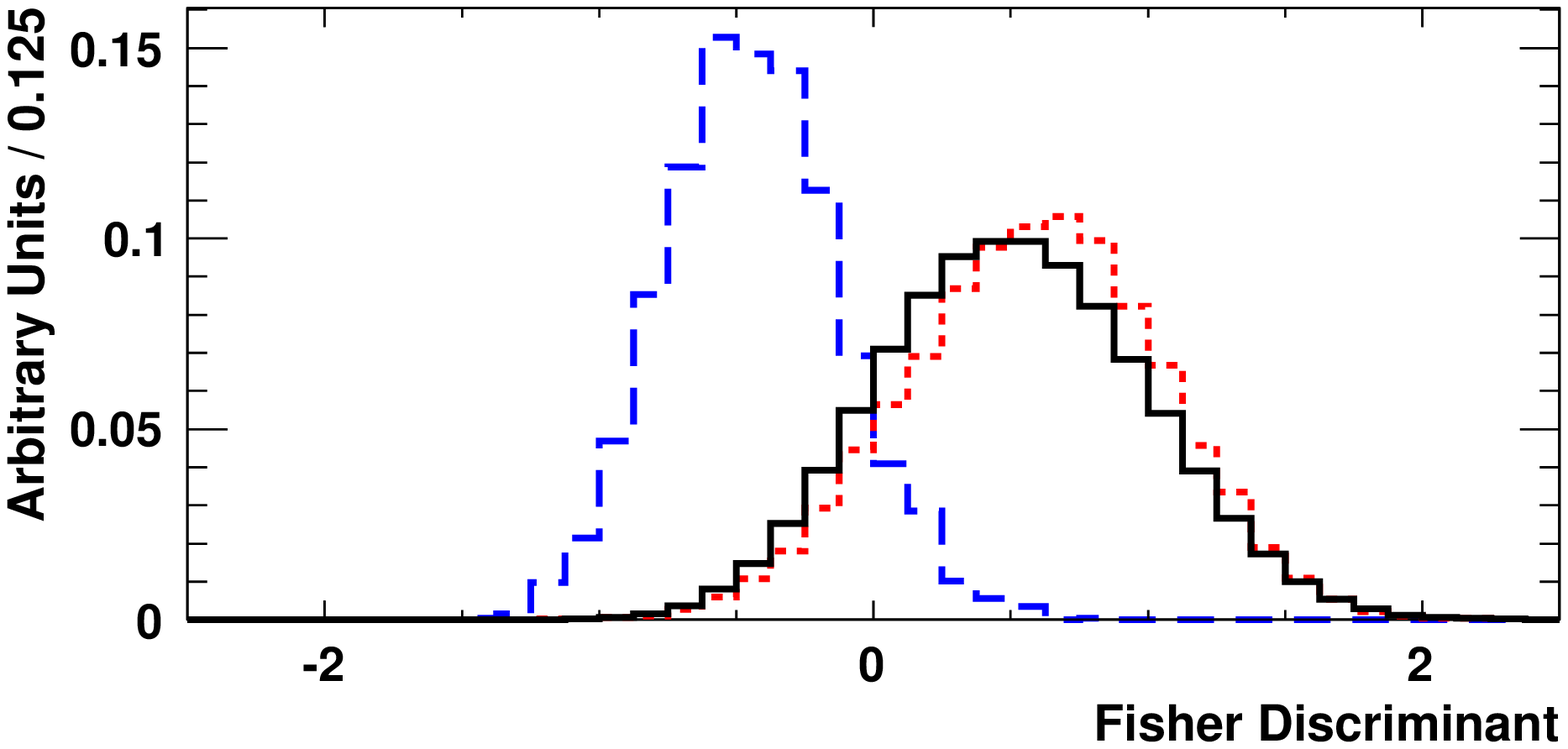}
         \caption{ 
                  Fisher discriminant, ${\cal F}$, distributions for
                  samples of $B\bar{B}$ MC (dashed blue), \dtomumu signal MC
                  (solid black), and continuum MC (dotted red).
                  The ${\cal F}$ distributions for \dtoee and \dtoemu
                  are similar to those of \dtomumu.
                 }
         \label{fig:fisher}
      \end{center}
   \end{figure}

   Figure~\ref{fig:fisher} shows distributions of the Fisher discriminant (${\cal F}$)
   for samples of \bbbar MC, \dtomumu signal MC, and continuum background MC.
   The separation between signal and \bbbar background distributions is large,
   while the signal and continuum background distributions are similar.
   For example, requiring ${\cal F}$ to be greater than 0
   removes about 90\%
   of the \bbbar background while keeping 85\% of the signal.
   The minimum ${\cal F}$ value is optimized for each
   signal channel as described below.

\begin{figure*}
      \begin{center}
         \includegraphics[width=0.31\linewidth]{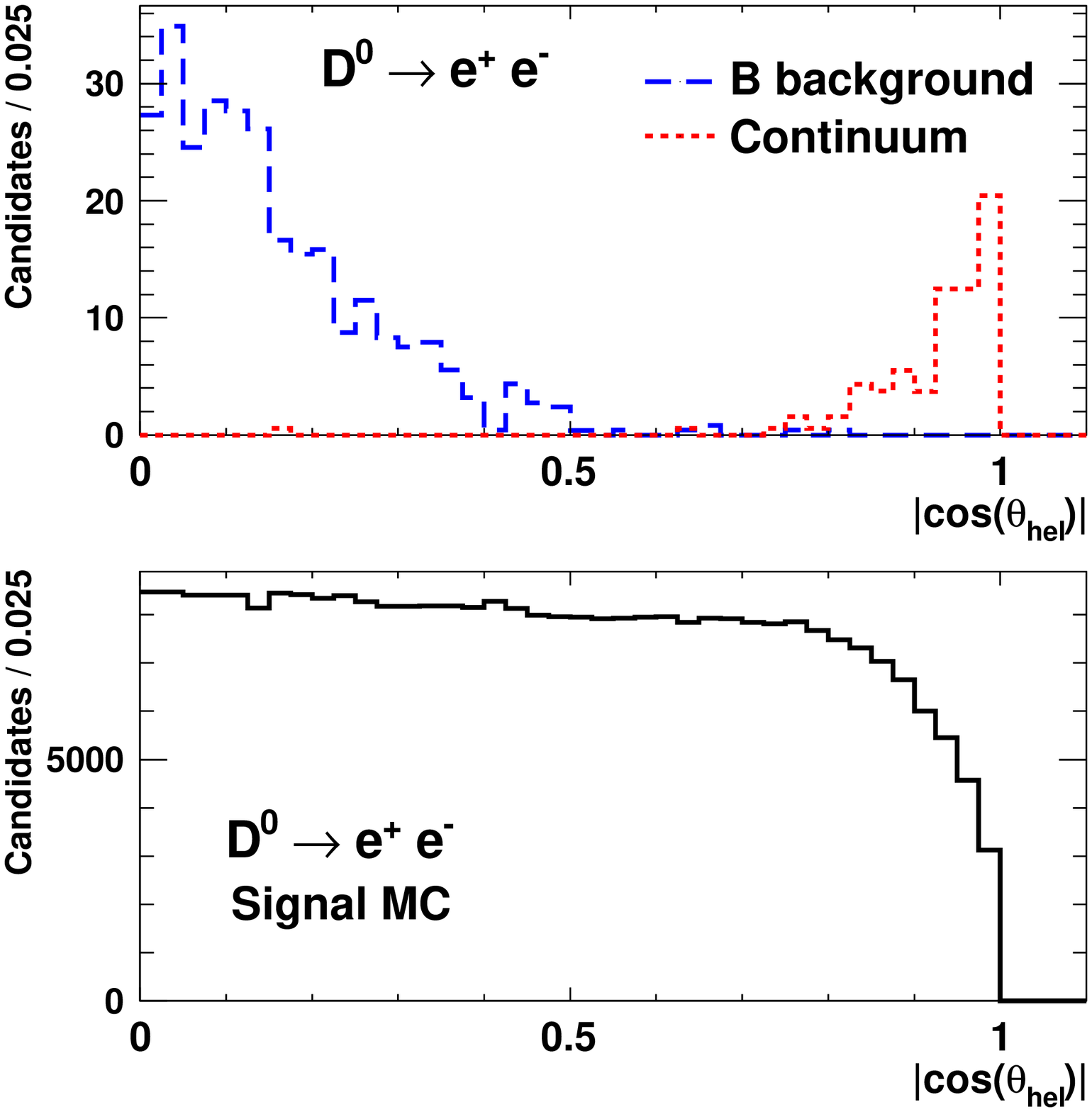}
         \includegraphics[width=0.31\linewidth]{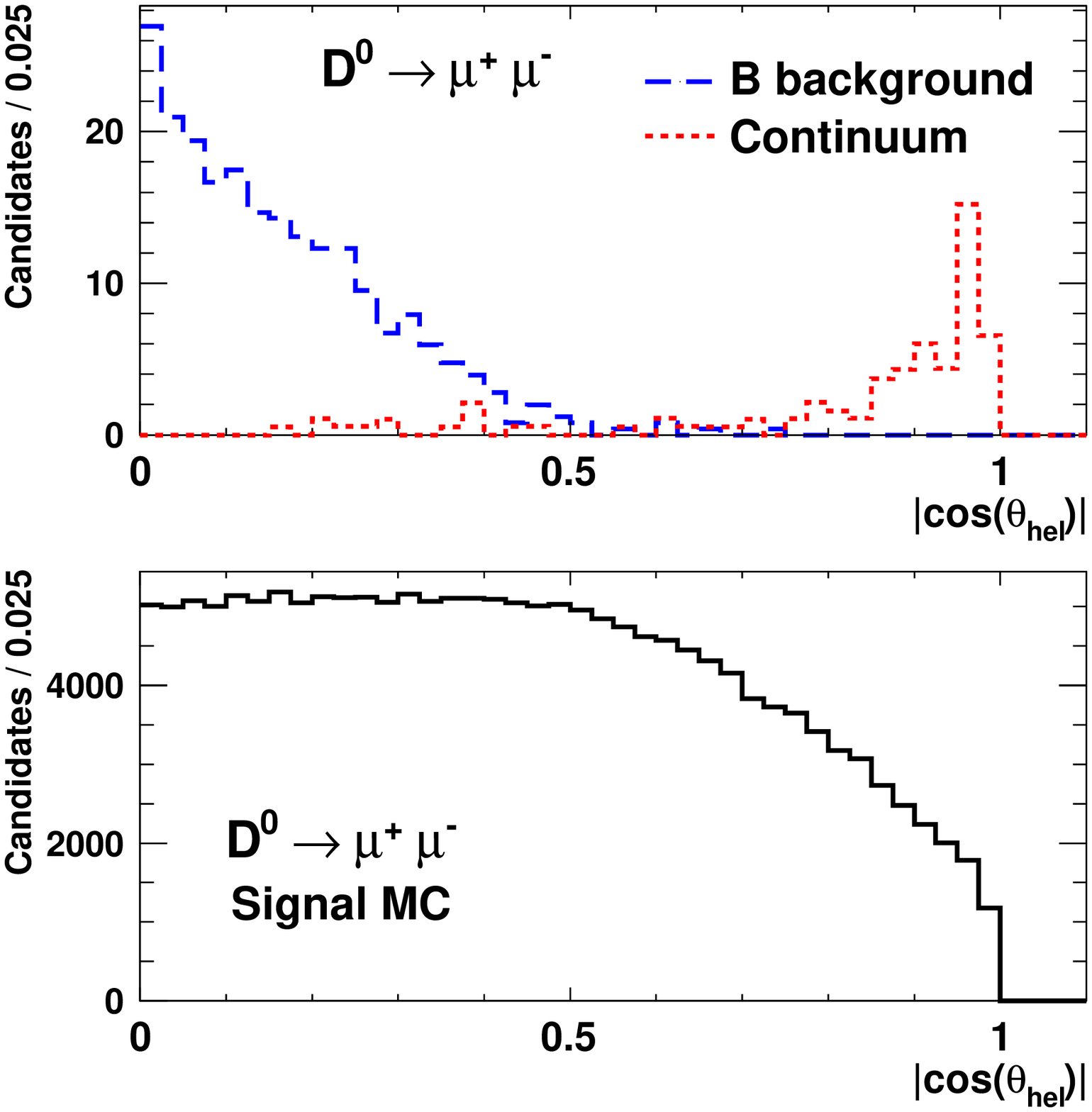}
         \includegraphics[width=0.31\linewidth]{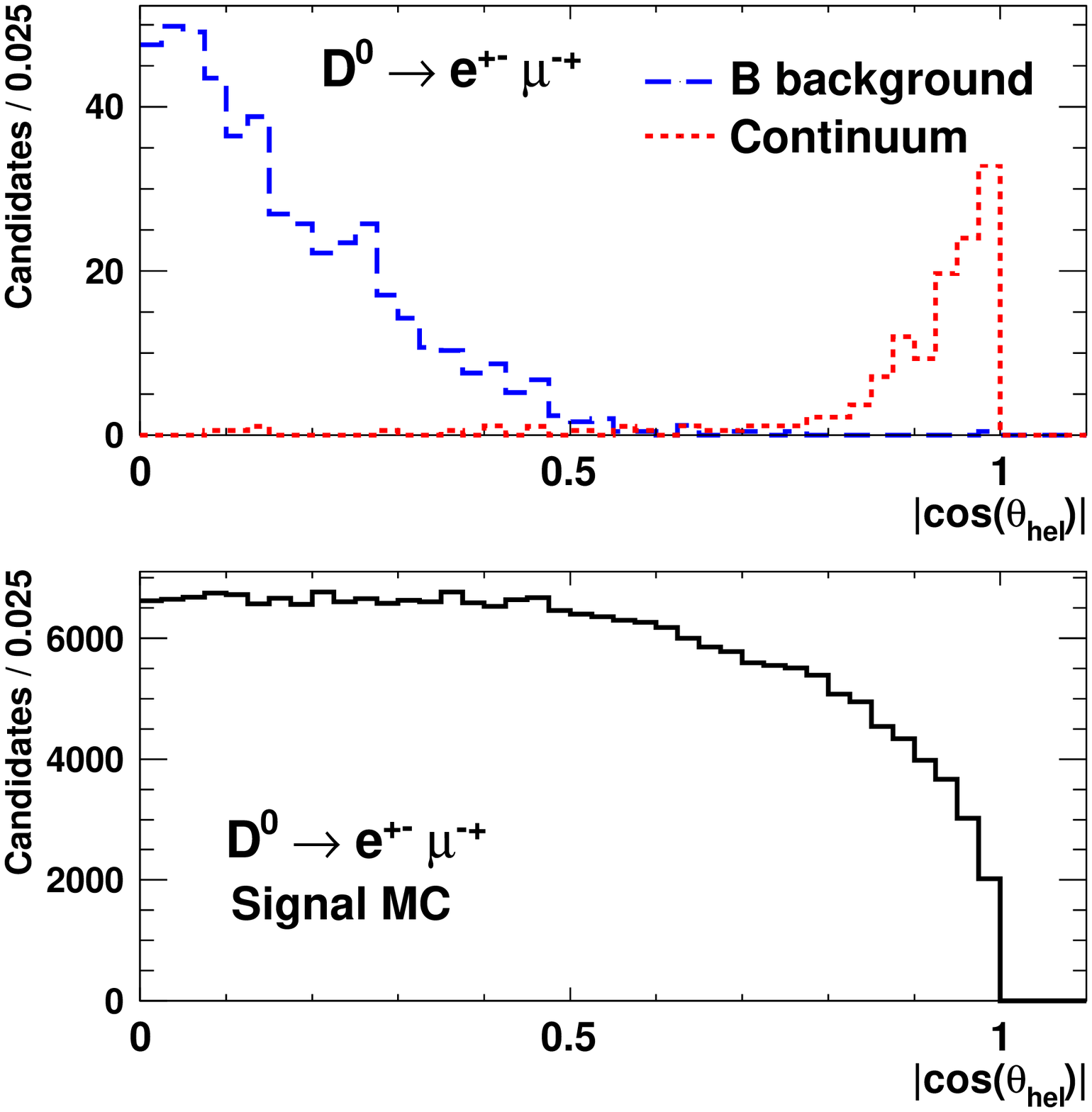}
         \caption{
                  Distributions of $|\cos(\theta_{\rm hel})|$ for the three signal channels:
                  \dtoee (left), \dtomumu (center), and \dtoemu (right).
                  The top distributions show Monte Carlo distributions for
                  the combinatorial \BB\ (dashed, blue) and continuum (dotted, red) backgrounds.
                  The bottom distributions show the signal Monte Carlo with
                  arbitrary normalization.
                 }
         \label{fig:coshel}
      \end{center}
   \end{figure*}

   We use the \coshel\ variable directly to remove continuum combinatorial background.
   Figure~\ref{fig:coshel} shows distributions of \coshel\, before making a minimum
   ${\cal F}$ requirement, for \bbbar background, continuum background, and signal.
   The drop-off for \coshel\ near $1.0$ in the signal distributions is caused by the
   selection and particle identification requirements.
   The \bbbar background peaks near zero, while the continuum background peaks
   sharply near one.

   The selection criteria for each signal channel were chosen
   to give the lowest expected signal branching fraction upper limit
   for the null hypothesis (a true branching fraction of zero)
   using the MC samples.
   The Fisher discriminant coefficients were determined before applying
   the \coshel, \dzero mass, and \deltam\ requirements.
   We then tested a total of 2700 configurations of \coshel, ${\cal F}$,
   \dzero mass, and \deltam\ criteria.
   Table~\ref{tab:optimal-cuts} summarizes
   the resulting best values for the maximum \coshel, minimum ${\cal F}$, \mdzero\
   signal window, and \deltam\ interval.

    \begin{table}
      \begin{center}
        \caption{ Selection criteria for the three \dtoll signal decay modes.
               The parameter in the last row is defined as
               $\delta \Delta m \equiv \Delta m - \Delta m_0$, where
               $\Delta m_0$ is the nominal $D^{*+}-D^0$ mass difference~\cite{PDG}.
             }
        \label{tab:optimal-cuts}
        \begin{tabular}{cccc}
          \hline\hline
            Parameter  & \ \ \ \ \ \ \ \ee   \ \ \ \ \ \
                       & \ \ \ \ \ \ \ \mumu \ \ \ \ \ \
                       & \ \ \ \ \ \emu  \ \      \\
            \hline
            \coshel     &  $<0.85$  &  $<0.90$  &  $<0.85$  \\
            ${\cal F}$  &  $>0.00$  &  $>-0.25$  &  $>0.00$  \\
            \mdzero (GeV) & $[1.815,1.890]$  & $[1.855,1.890]$  & $[1.845,1.890]$ \\
            $|\delta\Delta m|$ (MeV) & $<0.5$ & $<0.5$  & $<0.4$ \\
          \hline\hline
        \end{tabular}
      \end{center}
    \end{table}

    After the selection criteria in Table~\ref{tab:optimal-cuts} were determined,
    the data yields in the sideband region
    were compared to the expectations from Monte Carlo samples.
    The \dtomumu and \dtoemu data yields were consistent with the expectations
    from the Monte Carlo samples.
    However, the \dtoee sideband yield showed a substantial excess of events;
    90 events were observed when $5.5 \pm 1.6$ were expected.

    The excess of data sideband events over the expected background from
    Monte Carlo was investigated  and found to have several distinct features:
    low track multiplicity, continuum-like event shape characteristics,
    tracks consistent with electrons produced in photon conversions,
    low \dzero daughter track momenta, and undetected energy along the
    beam axis.
    We found that such events result from hard initial state radiation events
    or two-photon interaction processes that are not simulated in the continuum MC samples
    used in the analysis.
    The following selection criteria were added in order to remove such background contributions:
    \begin{itemize}
      \item  Events must have at least 5 tracks for the \dtoee channel and
             at least 4 tracks for the \dtomumu and \dtoemu channels.
      \item  Events can have at most 3 electron candidates.
      \item  The longitudinal boost of the event, reconstructed
             from all tracks and neutral clusters,
             along the high-energy beam direction $p_z/E$ in the CM frame
             must be greater than -0.5 for all three \dtoll channels.
      \item  For \dtomumu and \dtoemu candidates, the pion track from the $D^{*+}$
             decay and the leptons must be inconsistent with originating
             from a photon conversion.
    \end{itemize}
    The signal efficiencies for the \dtoee, \dtomumu, and \dtoemu channels
    for these additional criteria are 91.4\%, 99.3\%, and 96.8\%, respectively.
    The \dtoee sideband yield in the data with these criteria applied is
    reduced to 8 events where $4.5\pm 1.3$ are expected, based on the Monte Carlo samples.

   \subsection{ \boldmath Peaking \dtopipi background estimation }

   The amount of \dtopipi peaking background within the \mdzero signal window
   is estimated from data and
   calculated separately for each \dtoll channel using
   \begin{equation}
      N_{\pi\pi}^{BG} =
         \left( \sum_i N_{\pi\pi,i}^{NP}
            \cdot \langle p_{f,i}^+\rangle \langle p_{f,i}^-\rangle \right)
           \cdot \epsilon_{m(D^0)} \cdot G
   \end{equation}
   where the sum $i$ is over the six data-taking periods,
   $N_{\pi\pi,i}^{NP}$ is the number of \dtopipi events that
   pass all of the \dtoll selection criteria except for the
   lepton identification and \mdzero signal window requirements,
   $\langle p_{f,i}^+\rangle \langle p_{f,i}^-\rangle$ is the product of the
   average probability that the $\pi^+$ and the $\pi^-$ pass the lepton
   identification criteria,
   $\epsilon_{m(D^0)}$ is the efficiency for \dtopipi background to satisfy
   the \mdzero signal window requirement,
   and $G$ takes into account a positive correlation in
   the probability that the $\pi^+$ and the $\pi^-$ pass the muon identification
   criteria.
   The value of $\langle p_{f,i}^+\rangle$  $(\langle p_{f,i}^-\rangle)$ is
   measured using the ratio of
   the \dtopipi yield requiring that  the $\pi^+$  $(\pi^-)$ satisfy the
   lepton identification requirements
   to the \dtopipi yield with no lepton identification requirements applied.
   The $\langle p_{f,i}^+\rangle$ and $\langle p_{f,i}^-\rangle$ are measured
   separately for each of the six major data-taking periods due to the changing
   IFR performance with time.
       The values of $\langle p_{f,i}^+\rangle$ and $\langle p_{f,i}^-\rangle$
       vary between 0.5\% and 1.5\%.
   The probability that the $\pi^+$ and $\pi^-$ both pass the muon identification
   criteria is enhanced when the two tracks curve toward each other, instead of
   away from each other, in the plane perpendicular to the beam axis.
   We use $G = 1.19 \pm 0.05$ for the \dtomumu channel and
   $G = 1$ for the \dtoee and \dtoemu channels.
   The $G$ factor is measured using a high-statistics \dtokpi sample where
   the $K$ is required to have a signature in the IFR that matches that of
   a $\pi$ which passes the $\mu$ identification criteria.
   This is in good agreement with the MC estimate of the $G$
   factor value, $1.20 \pm 0.10$.

   \subsection{ Combinatorial background estimation }

  The combinatorial background is estimated by using the number of observed events
  in a sideband region and the expected ratio of events $R_{\rm cb}$ in the signal and
  sideband regions, determined from MC simulation.
  The sideband is above the signal region in the \dzero mass
  ([1.90, 2.05] GeV) in a wide \deltam window ([0.141, 0.149] GeV).
  We fit the \dzero mass and \deltam projections of the combinatorial background
  MC using $2^{\rm nd}$-order polynomials.
  A two-dimensional probability density function (PDF) is formed by multiplying
  the one-dimensional PDFs, assuming the variables are uncorrelated.
  The combinatorial background signal-to-sideband ratio $R_{\rm cb}$ is then computed from the
  ratio of the integrals of the two-dimensional PDF.

  \section{ Results }
\begin{figure*}
      \begin{center}
         \includegraphics[width=0.31\linewidth]{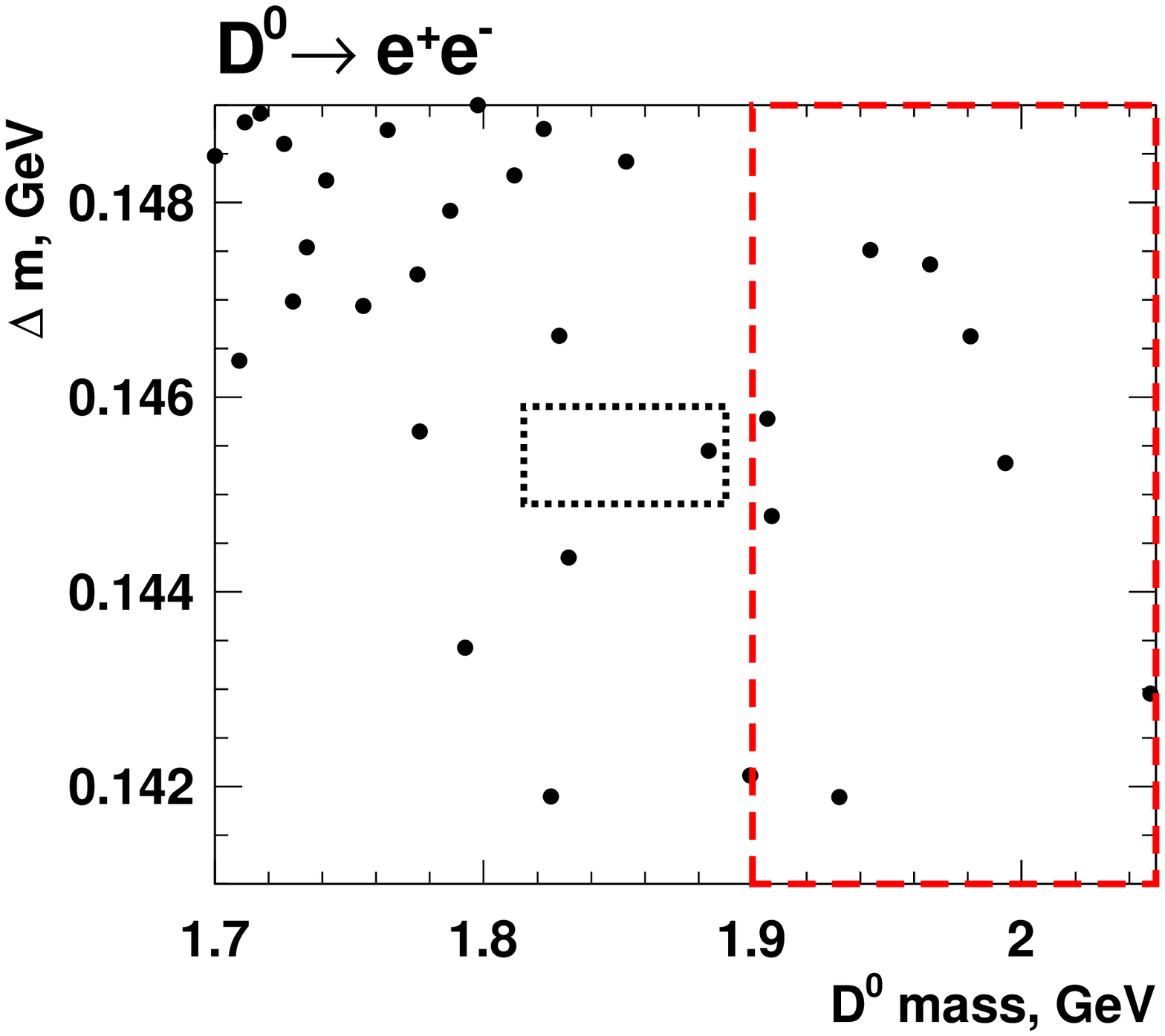}
         \includegraphics[width=0.31\linewidth]{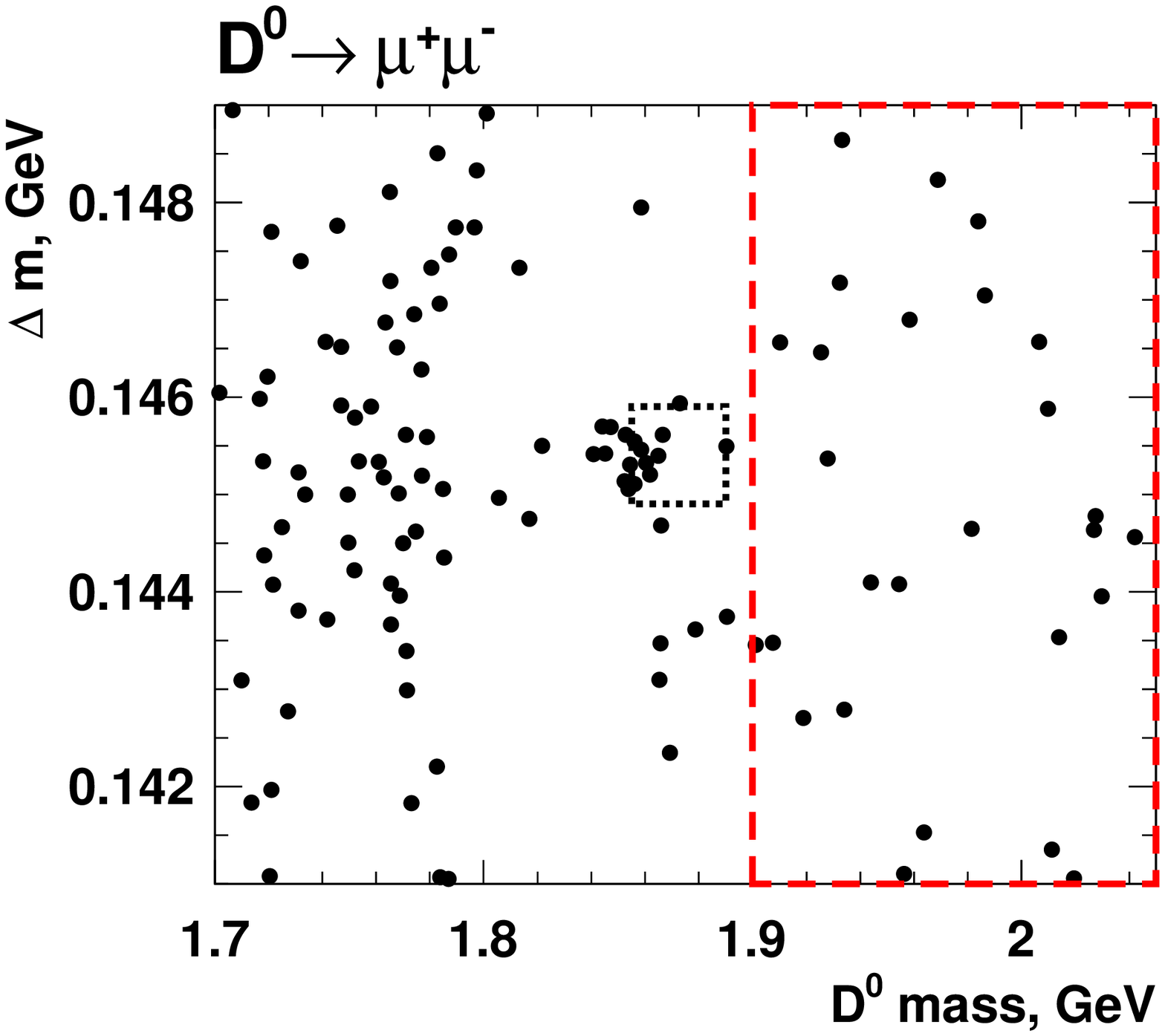}
         \includegraphics[width=0.31\linewidth]{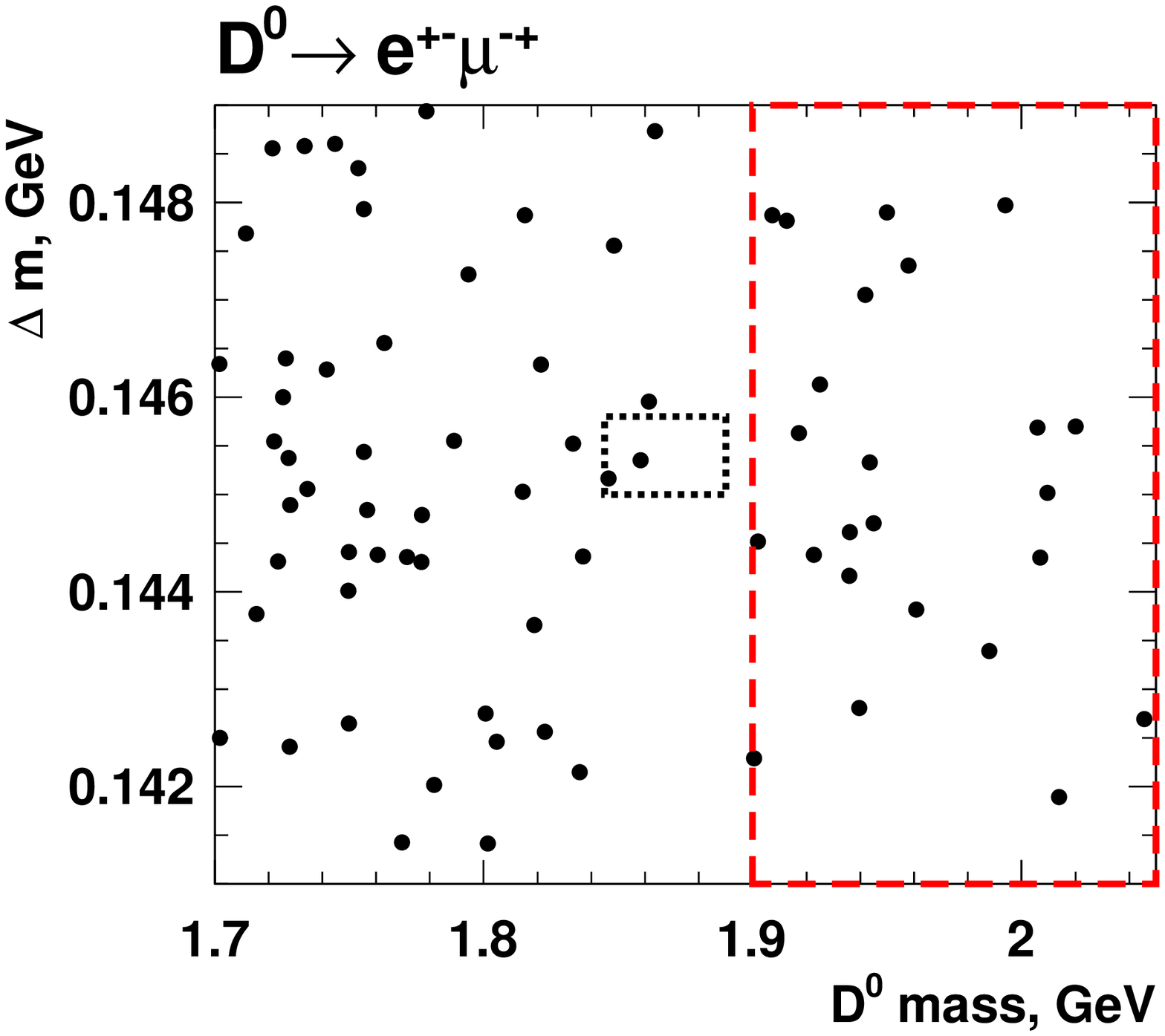}
         \includegraphics[width=0.31\linewidth]{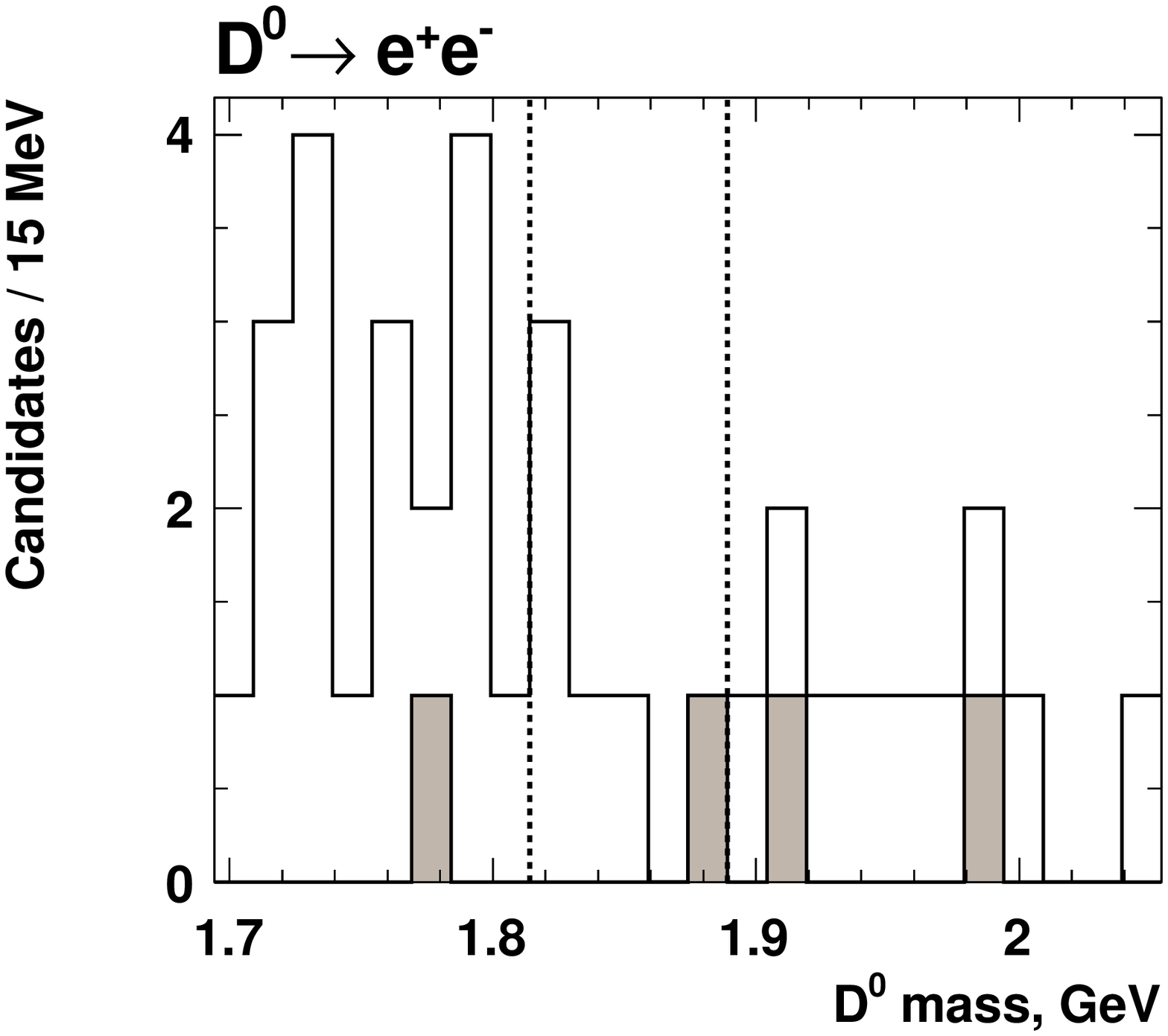}
         \includegraphics[width=0.31\linewidth]{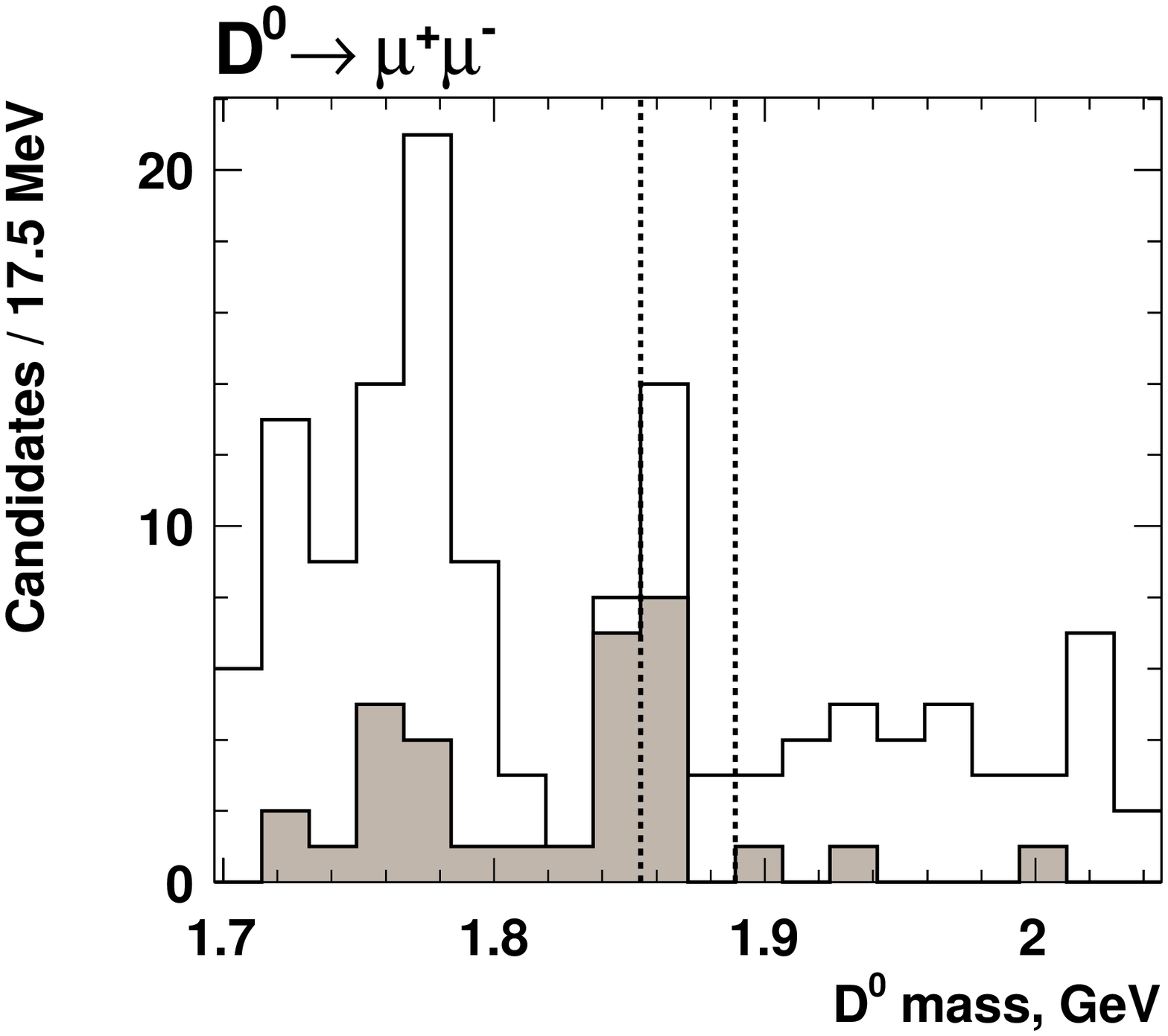}
         \includegraphics[width=0.31\linewidth]{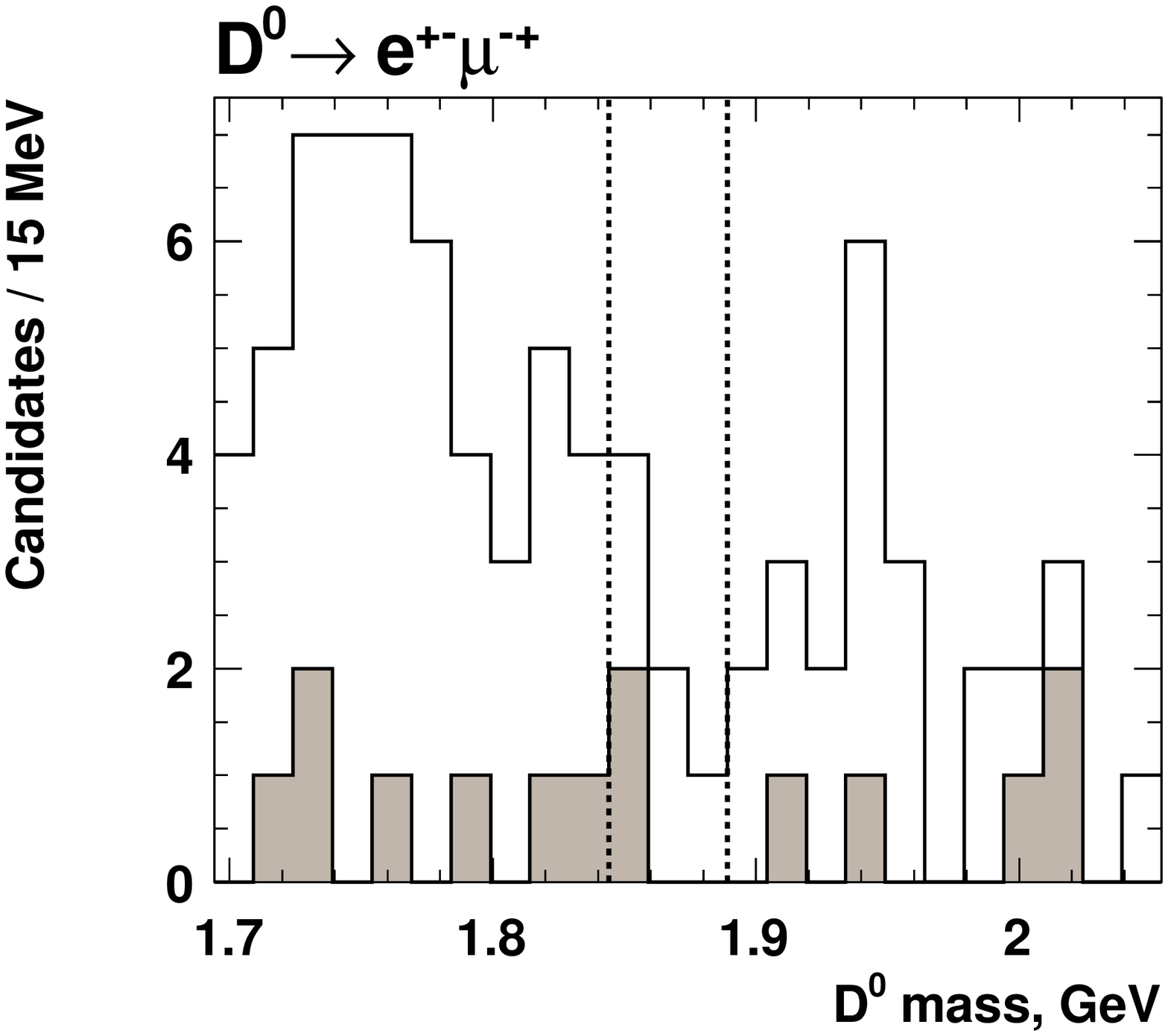}
         \includegraphics[width=0.31\linewidth]{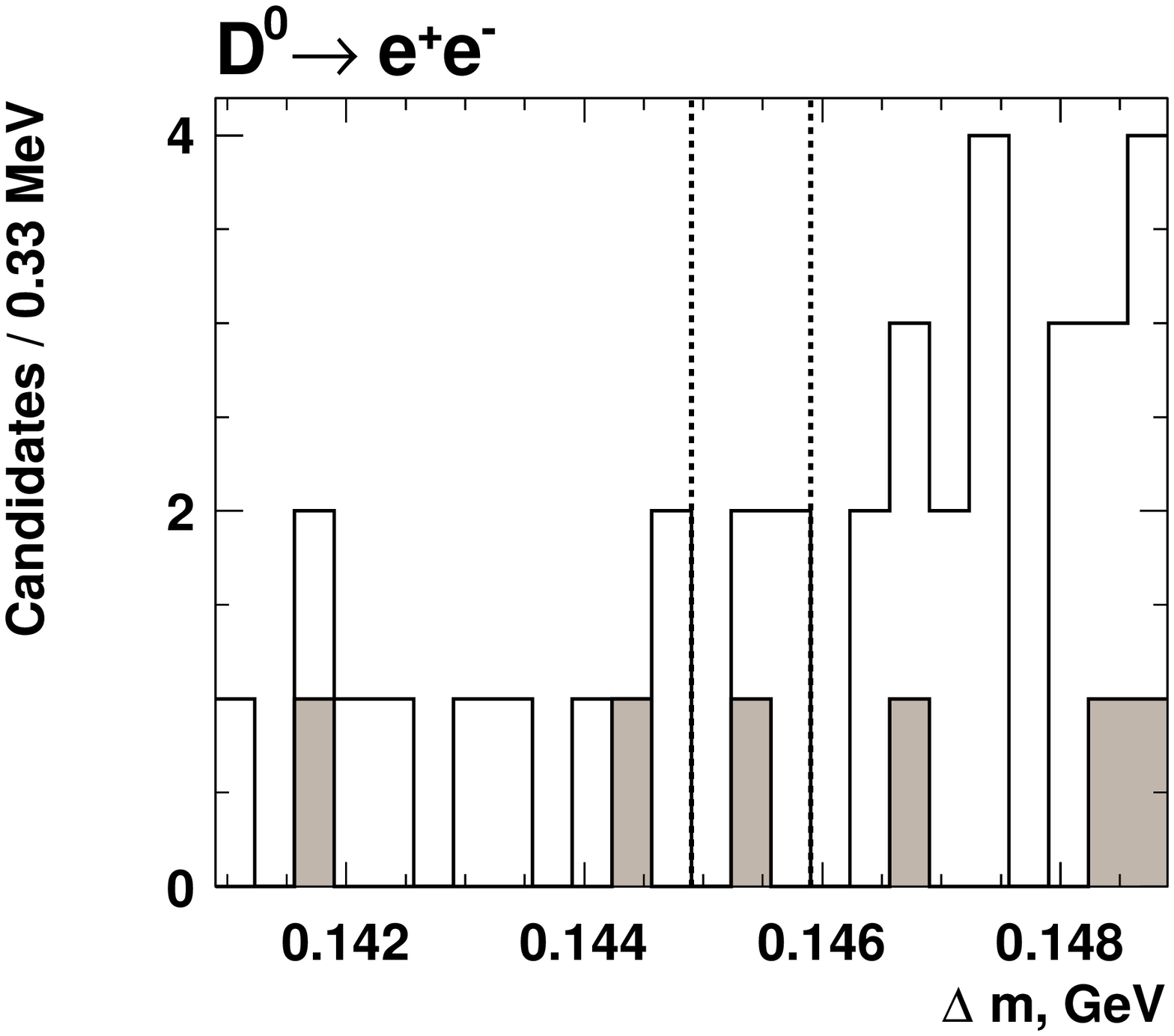}
         \includegraphics[width=0.31\linewidth]{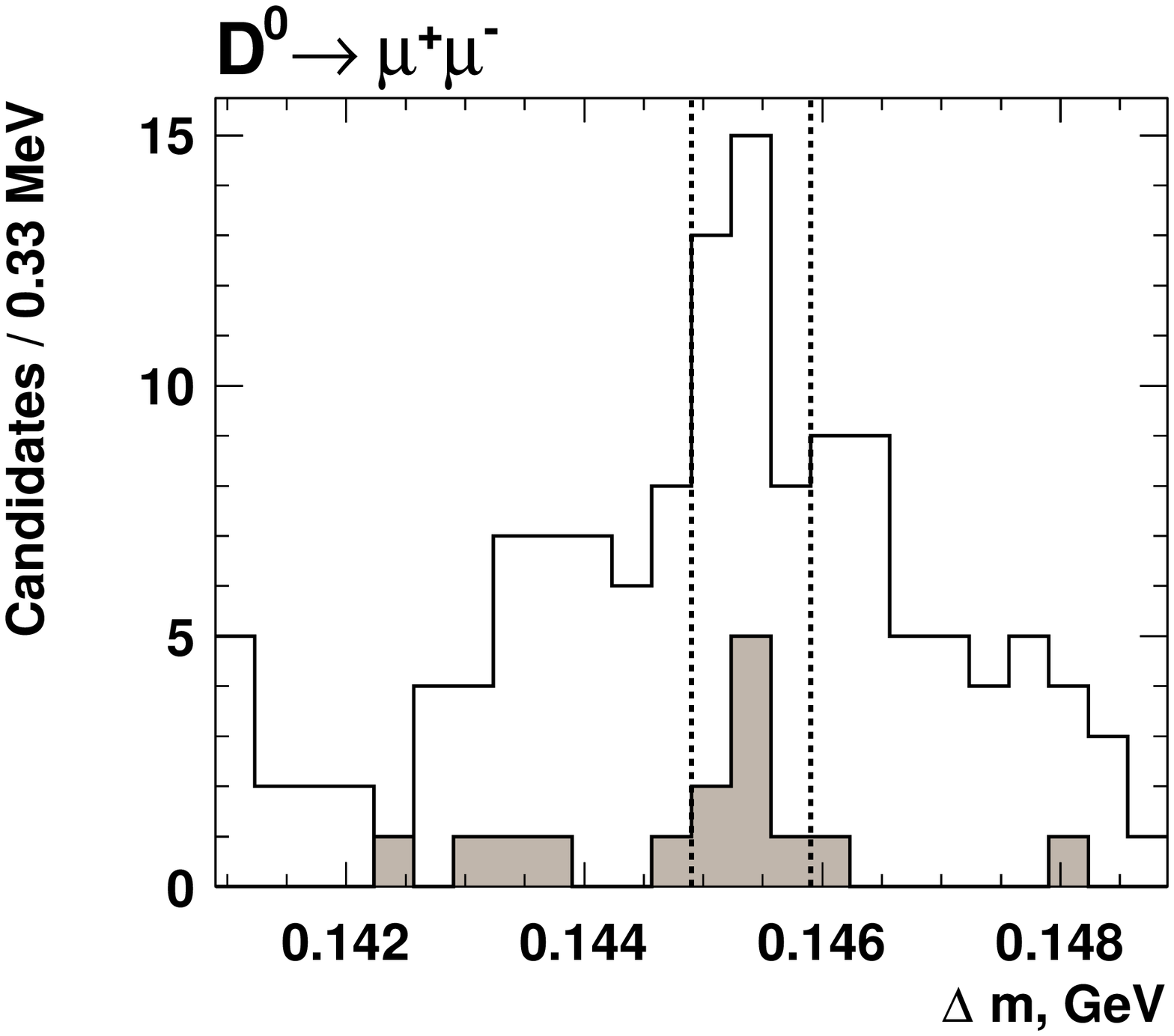}
         \includegraphics[width=0.31\linewidth]{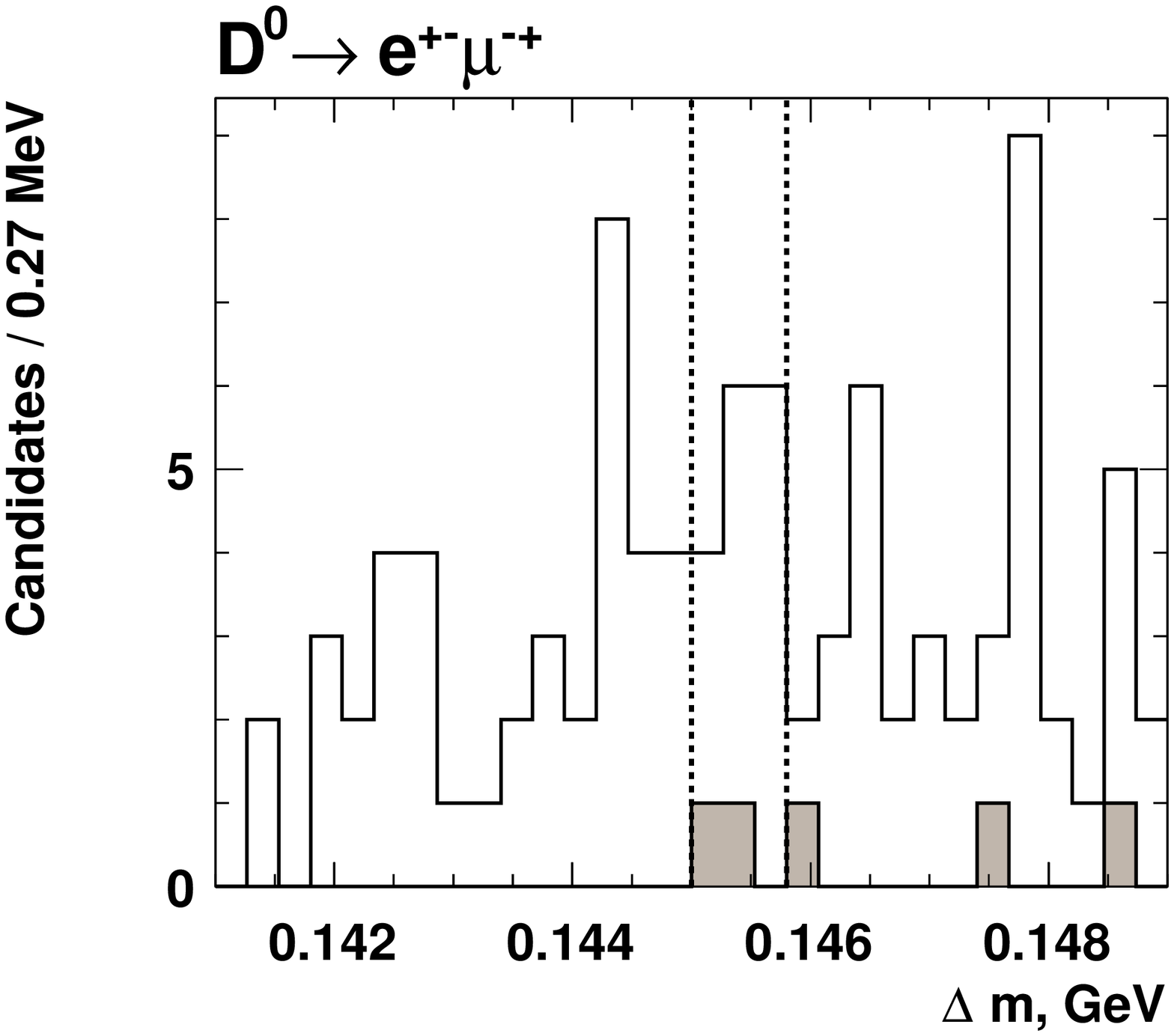}
         \caption{
                  Data distributions of $\Delta m$ vs
                  the reconstructed $D^0$ mass (top row)
                  and projections of the \dzero mass (middle row)
                  and $\Delta m$ (bottom row).
                  The columns contain the distributions for the \dtoee (left),
                  \dtomumu (center), and \dtoemu (right)
                  decay modes.
                  The shaded \dzero mass ($\Delta m$) distributions represent the
                  subset of events that fall in the $\Delta m$ (\dzero mass)
                  signal window.
                  In the top row, the dotted (black) box indicates the signal region
                  and the dashed (red) box indicates the sideband region.
                  In the middle and bottom rows, the vertical dotted black lines
                  indicate the boundaries of the signal region.
                 }
         \label{fig:dmvsd0m}
      \end{center}
   \end{figure*}

   The distribution of \deltam vs \dzero mass
   as well as projections of \deltam and the \dzero mass
   for the data
   events for the three signal channels are shown in Fig.~\ref{fig:dmvsd0m}.
   Peaks from \dtokpi and \dtopipi are visible at 1.77~GeV and
   1.85~GeV in the \dzero mass distribution for \dtomumu candidates.
   We observe 1, 8, 2 events in the \dtoee, \dtomumu, and \dtoemu
   signal regions, respectively.


   \subsection{ \boldmath \dtoll Branching fractions }

   The yield of \dtopipi decays in the $\pi\pi$ control sample,
   selected with the same ${\cal F}$ and $|\cos \theta_{\rm hel}|$ criteria
   for each \dtoll signal mode (see Table~\ref{tab:optimal-cuts}), is used to normalize the
   \dtoll signal branching fraction.
   For each \dtoll signal channel, the \dtopipi yield $N_{\pi\pi}^{\rm fit}$ is determined
   by fitting the \dzero mass spectrum of the \dtopipi control sample in the range [1.7, 2.0] GeV.
   The fit has three components: \dtopipi, \dtokpi, and combinatorial background.
   The PDF for the \dtopipi component is the sum of a Crystal Ball function and
   two Gaussians.
   The Crystal Ball function is a Gaussian modified to have an extended, power-law
   tail on the low side~\cite{CB-function}.
   The PDF for the \dtokpi component is the sum of a Crystal Ball function and an
   exponential function.
   The combinatorial background PDF is an exponential function.

   The \dtoll branching fraction is given by
   \begin{equation}
      {\cal B}_{\ell\ell} = \left(\frac{N_{\ell\ell}}{N_{\pi\pi}^{\rm fit}}\right)
              \ \left(\frac{\epsilon_{\pi\pi}}{\epsilon_{\ell\ell}}\right)
              \ {\cal B}_{\pi\pi}
              \ = \ S_{\ell\ell} \ \cdot \ N_{\ell\ell}
   \end{equation}
   where  $N_{\ell\ell}$ is the number of \dtoll signal candidates,
   $N_{\pi\pi}^{\rm fit}$ is the number of \dtopipi candidates from the fit,
   $\epsilon_{\pi\pi}$ and $\epsilon_{\ell\ell}$ are the efficiencies for the
   corresponding decay modes, ${\cal B}_{\pi\pi} = (1.400 \pm 0.026) \times 10^{-3}$
   is the \dtopipi branching fraction~\cite{PDG}, and $S_{\ell\ell}$ is
   defined by
   \begin{equation}
     \label{eq:S}
      S_{\ell\ell} \ \equiv \ \frac{ {\cal B}_{\pi\pi} }
                                     {    N_{\pi\pi}^{\rm fit} }
                           \frac{ \epsilon_{\pi\pi} }{ \epsilon_{\ell\ell} }.
   \end{equation}
   The expected observed number of events in the signal region is given by
   \begin{equation}
      N_{\rm obs} \ = \ {\cal B}_{\ell\ell} / S_{\ell\ell} + N_{BG}.
   \end{equation}
   The uncertainties on $S_{\ell\ell}$ and $N_{BG}$ are incorporated into
   a likelihood function by convolving a Poisson PDF in $N_{\rm obs}$
   with Gaussian PDFs in $S_{\ell\ell}$ and $N_{BG}$.
   We determine 90\% confidence level intervals using the likelihood
   ratio ordering principle of Feldman and Cousins~\cite{feldman}
   to construct the confidence belts.
   The estimated branching fractions and one standard deviation uncertainties
   are determined from the values of ${\cal B}_{\ell\ell}$ that maximize
   the likelihood and give a change of 0.5 in the log likelihood
   relative to the maximum, respectively.

   \subsection{ Systematic uncertainties }

   Table~\ref{tab:syst} summarizes the systematic uncertainties.
   Several of the uncertainties in $\epsilon_{\pi\pi} / \epsilon_{\ell\ell}$ cancel,
   including tracking efficiency for the \dzero daughters, slow pion efficiency, and the
   efficiencies of the ${\cal F}$ and \dzero momentum requirements.
   The uncertainty on $\epsilon_{\pi\pi}/\epsilon_{\ell\ell}$ due to particle
   identification is 4\%.
   Bremsstrahlung creates a low-side tail in the \dzero mass distributions for
   the \dtoee and \dtoemu decay modes.
   The uncertainty $\epsilon_{\ell\ell}$ due to the modeling of this tail
   is 3\% for \dtoee and 2\% for \dtoemu.
   The Crystal Ball shape parameters that describe the low-side tail of the
   \dzero mass distribution were varied, leading to an uncertainty of 1.1\% to 1.3\%
   on $N^{\rm fit}_{\pi\pi}$.
   We use the world average for the \dtopipi branching fraction~\cite{PDG},
   which has an uncertainty of 1.9\%.
   We combine the above relative uncertainties in quadrature resulting in
   4.6\% to 5.4\% systematic uncertainties on $S_{\ell\ell}$.

   The \dzero mass range for the fit used to determine the combinatorial background
   PDF was varied from [1.70, 2.05] GeV to [1.80, 2.05] GeV.
   The difference in the resulting signal-to-sideband ratio $R_{\rm cb}$
   is taken as a systematic uncertainty.
   The pion misidentification probabilities for $e$ and $\mu$ measured in data
   are in good agreement with the MC simulation.
   We use the larger of either the difference between the data and the MC
   or the statistical uncertainty on the MC misidentification probabilities
   as a systematic uncertainty.
   For the \dtomumu decay mode, we take the uncertainty on the MC
   estimate for the $G$ factor of 8\% as a systematic uncertainty
   on the $G$ estimate from the \dtokpi data control sample.

    \begin{table*}
      \begin{center}
    \caption{ Systematic uncertainties.
        The uncertainty on $S_{\ell\ell}$ results from the uncertainties
        on $\epsilon_{\pi\pi}/\epsilon_{\ell\ell}$,
        $N^{\rm fit}_{\pi\pi}$, and ${\cal B}_{\pi\pi}$ added in quadrature.
        The systematic uncertainty on the overall background $N_{BG}$ is obtained
        from the uncertainties on $N^{BG}_{\pi\pi}$ and $N_{\rm cb}$ added in
        quadrature.
          }
     \label{tab:syst}
   \begin{tabular}{lccc}
   \hline\hline
       & $D^0\to e^+e^-$
       & $D^0\to \mu^+\mu^-$
       & $D^0\to e^\pm\mu^\mp$ \smtvs
       \\
   \hline\hline
   $\epsilon_{\pi\pi}/\epsilon_{\ell\ell}$, particle ID  &  4\%  &  4\%  &  4\%  \smtvs  \\
   \hline
   $\epsilon_{\pi\pi}/\epsilon_{\ell\ell}$, Bremsstrahlung \ \ \ \  &  3\%  &  ---  &  2\%  \smtvs  \\
   \hline
   $N^{\rm fit}_{\pi\pi}$  &  1.2\%  &  1.3\%  &  1.1\%  \smtvs \\
   \hline
   ${\cal B}_{\pi\pi}$  &  1.9\%  &  1.9\%  &  1.9\%  \smtvs \\
   \hline
   $S_{\ell\ell}$  &  5.4\%  &  4.6\%  &  5.0\%  \smtvs \\
   \hline \hline
   $N^{BG}_{\pi\pi}$  & \ \ \ \  11\% (0.004 events) \ \ \ \
                      & \ \ \ \  16\% (0.43 events)  \ \ \ \
                      & \ \ \ \   5\% (0.02 events)  \ \ \ \ \smtvs \\
   \hline
   $N_{\rm cb}$, $R_{\rm cb}$  &  36\% (0.35 events)  &  20\% (0.25 events)  &  19\% (0.20 events)  \smtvs \\
   \hline
   $N_{BG}$  &  0.35 events  &  0.50 events  &  0.20 events \\
   \hline\hline
     \end{tabular}
     \end{center}
     \end{table*}


   \subsection{ Branching Fraction Results }


    \begin{table*}
      %
      %
      \begin{center}
    \caption{ Results for the observed event yields ($N_{\rm obs}$),
              estimated background ($N_{BG}$),
              and signal branching fractions (${\cal B}_{\ell\ell}$).
              The first uncertainty is statistical and the second
              systematic.
              $N_{SB}$ is the observed number of events in the sideband,
              $R_{\rm cb}$ is the signal-to-sideband ratio for combinatorial background,
              $N_{\rm cb}$ and $N_{\pi\pi}^{BG}$ are the estimated combinatorial and
              \dtopipi backgrounds in the signal region,
              $N^{\rm fit}_{\pi\pi}$ is the fitted yield in the \dtopipi control sample,
              $\epsilon_{\pi\pi}$ and $\epsilon_{\ell\ell}$ are the $\pi\pi$ control sample and
              signal selection efficiencies, determined from Monte Carlo samples, which
              have negligible statistical uncertainties.
              The systematic uncertainty on $\epsilon_{\pi\pi}/\epsilon_{\ell\ell}$
              is included in the systematic uncertainty on $S_{\ell\ell}$, which
              is defined in Eqn.~(\ref{eq:S}).
          }
     \label{tab:results}
   \begin{tabular}{lccc}
   \hline\hline
       & $D^0\to e^+e^-$
       & $D^0\to \mu^+\mu^-$
       & $D^0\to e^\pm\mu^\mp$ \smtvs
       \\
   \hline\hline
    $N_{SB}$         & 8   & 27   & 24   \smtvs \\
   \hline
   $R_{\rm cb}$        &   \ \ \ \   $0.121 \pm 0.023 \pm 0.044$  \ \ \ \  
                       &   \ \ \ \   $0.046 \pm 0.005 \pm 0.009$  \ \ \ \  
                       &   \ \ \ \   $0.042 \pm 0.006 \pm 0.008$  \ \ \ \   \smtvs
                       \\
   \hline
   $N_{\rm cb}$        &   \ \ \ \  $0.97 \pm 0.39 \pm 0.35$  \ \ \ \
                       &   \ \ \ \  $1.24 \pm 0.27 \pm 0.25$  \ \ \ \
                       &   \ \ \ \  $1.00 \pm 0.25 \pm 0.20$  \ \ \ \ \smtvs
                         \\
   \hline
   $N_{\pi\pi}^{BG}$  &  $0.037 \pm 0.012 \pm 0.004$
                      &  $2.64  \pm 0.22  \pm 0.43 $
                      &  $0.42  \pm 0.08  \pm 0.02 $ \smtvs
                      \\
   \hline
   $N_{BG}$         &  $1.01 \pm 0.39 \pm 0.35$
                    &  $3.88 \pm 0.35 \pm 0.50$
                    &  $1.42 \pm 0.26 \pm 0.20$ \smtvs
                    \\
   \hline\hline
   $N^{\rm fit}_{\pi\pi}$  &  $39930 \pm 210 \pm 490$
                           &  $51800 \pm 240 \pm 660$
                           &  $39840 \pm 210 \pm 430$ \smtvs
                           \\
   \hline
   $\epsilon_{\pi\pi}$    &  14.4\%  &  18.7\%  &  14.6\%  \smtvs\\
   \hline
   $\epsilon_{\ell\ell}$  &  9.48\%  &  6.29\%  &  6.97\%  \smtvs \\
   \hline
   $S_{\ell\ell} \ (\times 10^{-9})$  &  $53.4 \pm 0.2 \pm 2.9$
                                      &  $80.6 \pm 0.4 \pm 3.7$
                                      &  $73.9 \pm 0.4 \pm 3.7$ \smtvs
                   \\
    \hline\hline
    $N_{\rm obs}$        &  1   &  8   &  2  \smtvs \\
    \hline
    ${\cal B}_{\ell\ell} \ (\times 10^{-7})$  &  $0.1\ ^{+0.7}_{-0.4}$  &  $3.3\ ^{+2.6}_{-2.0}$  &  $0.5\ ^{+1.3}_{-0.9}$  \smtvs \\
    \hline
    ${\cal B}_{\ell\ell} \ (\times 10^{-7})$ 90\% C.I. \ \ \ \ &  $<1.7$ & $[0.6,8.1]$ &  $<3.3$  \smtvs \\
   \hline\hline
   \end{tabular}
   \end{center}
    \end{table*}


   Table~\ref{tab:results} presents the results, where $N_{SB}$ is the number of
   events in the upper sideband, $N_{\rm cb}$ is the expected number of combinatorial background
   events in the signal window, $N_{\pi\pi}^{BG}$ is the number of events from the \dtopipi
   peaking background, and $N_{BG}$ (data) is the expected number of total background events
   in the data.

   For the \dtoee and \dtoemu channels, the event yield in the signal region
   is consistent with background only.
   We observe 1 and 2 events with expected backgrounds
   of $1.0 \pm 0.5$ and $1.4 \pm 0.3$ events for the \dtoee and \dtoemu channels, respectively.
   The 90\% confidence interval upper limits for the branching fractions are
   $<1.7\times 10^{-7}$ for \dtoee and $<3.3 \times 10^{-7}$ for \dtoemu.

   For the \dtomumu channel, we observe 8 events in the signal region,
   where we expect $3.9 \pm 0.6$ background events.
   There is a cluster of  of \dtomumu candidate events in Fig.~\ref{fig:dmvsd0m} just above
   and below the lower \dzero mass edge of the signal region, where the \dtopipi background
   is expected.
   We expect $7.5 \pm 0.8$ \dtopipi events in the entire [1.7, 2.05] GeV \dzero mass range,
   with 93\% of these events falling within the narrower [1.830,1.875] GeV range.
   The combinatorial background in the [1.830,1.875] GeV \dzero mass interval is expected to
   be $1.8 \pm 0.6$ events, giving a total expected background of $8.8 \pm 1.1$ events.
   In this interval, we observe 15 events.
   The probability of observing 15 or more events when $8.8 \pm 1.1$ events are expected is 4.6\%,
   which corresponds to a 1.7 standard deviation upward fluctuation from the mean
   for a Gaussian distribution (i.e. $({\rm Erf}(1.7/\sqrt{2})+1)/2 = 1 - 0.046$).
   The probability of observing 8 events when $3.9 \pm 0.6$ events are expected is 5.4\%.
   We conclude that the excess over the expected background is not statistically significant.
   The Feldman-Cousins method results in a two-sided 90\% confidence interval for the
   \dtomumu branching fraction of $[0.6, 8.1]\times 10^{-7}$.


   In summary, we have searched for the leptonic charm decays \dtoee, \dtomumu, and \dtoemu
   using 468~fb$^{-1}$ of integrated luminosity recorded by the \babar\ experiment.
   We find no statistically significant excess over the expected background.
   These results supersede our previous results~\cite{prev-babar} and are
   consistent with the results of the Belle experiment~\cite{belle}, which has
   set 90\% confidence level upper limits of
   $<0.79 \times 10^{-7}$,
   $<1.4 \times 10^{-7}$,
   and $<2.6 \times 10^{-7}$,
   for the \dtoee, \dtomumu, and \dtoemu branching fractions, respectively.
   The LHCb experiment has recently presented preliminary search results~\cite{lhcb} for
   \dtomumu, where they find no evidence for this decay and set an upper limit
   on the branching fraction of $< 1.3 \times 10^{-8}$ at 95\% C.L.

   \section*{ Acknowledgments }

   \par
   We are grateful for the 
extraordinary contributions of our \pep2\ colleagues in
achieving the excellent luminosity and machine conditions
that have made this work possible.
The success of this project also relies critically on the 
expertise and dedication of the computing organizations that 
support \babar.
The collaborating institutions wish to thank 
SLAC for its support and the kind hospitality extended to them. 
This work is supported by the
US Department of Energy
and National Science Foundation, the
Natural Sciences and Engineering Research Council (Canada),
the Commissariat \`a l'Energie Atomique and
Institut National de Physique Nucl\'eaire et de Physique des Particules
(France), the
Bundesministerium f\"ur Bildung und Forschung and
Deutsche Forschungsgemeinschaft
(Germany), the
Istituto Nazionale di Fisica Nucleare (Italy),
the Foundation for Fundamental Research on Matter (The Netherlands),
the Research Council of Norway, the
Ministry of Education and Science of the Russian Federation, 
Ministerio de Ciencia e Innovaci\'on (Spain), and the
Science and Technology Facilities Council (United Kingdom).
Individuals have received support from 
the Marie-Curie IEF program (European Union) and the A. P. Sloan Foundation (USA).



\begin{thebibliography}{99}
\bibitem{chrgconj} The use of charge conjugate processes is implied unless explicitly stated otherwise.
\end{thebibliography}
\end{document}